\documentclass[12pt,preprint]{aastex}

\def \arcsec{\hbox{$^{\prime\prime}$}}

\newcommand{\citeN}[1]{\citeauthor{#1} (\citeyear{#1})}
\newcommand{\citeNP}[1]{\citeauthor{#1} \citeyear{#1}}

\newcommand{\FeI}{\ion{Fe}{1}}
\newcommand{\OI}{\ion{O}{1}}
\newcommand{\NiI}{\ion{Ni}{1}}

\shortauthors{Centeno \& Socas-Navarro}

\begin{document}

\title{A new approach to the solar oxygen abundance problem}

\author{R. Centeno}
   	\affil{High Altitude Observatory, NCAR, 3080 Center Green Dr, Boulder, CO 80301, USA}
	\email{rce@ucar.edu}

\author{H. Socas-Navarro}
   	\affil{High Altitude Observatory, NCAR\altaffilmark{1},
	3080 Center Green Dr, Boulder, CO 80301, USA}
        \affil{Instituto de Astrof\'\i sica de Canarias, Avda V\'\i
          a L\' actea S/N, La Laguna 38205, Tenerife, Spain}
	\email{navarro@ucar.edu}

\altaffiltext{1}{The National Center
	for Atmospheric Research (NCAR) is sponsored by the National Science
	Foundation.}

\newcommand{\EO}{$\epsilon_{\rm O}$}
\newcommand{\lgEO}{${\rm log}\epsilon_{\rm O}$}
\newcommand{\EFe}{$\epsilon_{\rm Fe}$}
\newcommand{\lgEFe}{${\rm log}\epsilon_{\rm Fe}$}
\newcommand{\ENi}{$\epsilon_{\rm Ni}$}
\newcommand{\lgENi}{${\rm log}\epsilon_{\rm Ni}$}


\begin{abstract}
In this work we present new data that sets strong constraints on the
solar oxygen abundance. Our approach, based on the analysis of
spectro-polarimetric observations, is almost model-independent and
therefore extremely robust. The asymmetry of the Stokes~$V$ profile of
the 6300 \AA\, [\OI] and \NiI\, blend is used as an indicator of the
relative abundances of these two elements.  The peculiar shape of the
profile requires a value of \EO=730$\pm$100 ppm~(parts per million),
or \lgEO=8.86$\pm$0.07 in the logarithmic scale commonly used in
Astrophysics. The uncertainty range includes the model dependence as
well as uncertainties in the oscillator strenghts of the lines.  We
emphasize that the very low degree of model dependence in our analysis
makes it very reliable compared to traditional determinations.
\end{abstract}
   

\section{Introduction}

Oxygen is the third most abundant chemical element in the Universe,
after Hydrogen and Helium, and the one that is almost exclusively 
produced by nuclear fusion in stellar interiors. Its abundance in the
Sun was thought to be well established since the 1980s (\EO=850$\pm$80
parts per million, relative to hydrogen, ppm; \citeNP{anders-grevesse89}; more
recently \EO=680$\pm$100 \citeNP{grevesse-sauval98})\footnote{ In this
  paper we use a linear abundance scale, given in ppm, because it is a
  more convenient unit in our approach. Some times, however, the
  traditional Astrophysical logarithmic scale is given to allow for
  easy comparison with previous work (${\rm log}\epsilon=12+\log_{10}
  [\epsilon \times 10^{-6}]$).}.  
However, a recent work (\citeNP{asplundetal04}) using a
new 3D hydrodynamical model of the solar atmosphere (as well as
updated atomic and molecular data) recommends a revision of the O
abundance to a lower value of \EO=457$\pm$50 ppm (\lgEO=8.66 $\pm$
0.05). The revised solar composition is said to fit better within its galactic
environment but it also creates a serious problem, namely it ruins the
exceptionally good agreement between various predictions of solar interior
models and properties inferred from helioseismology (see, e.g., \citeNP{basu-antia2008}).

The chemical composition of celestial bodies is not a directly
measurable quantity. It is deduced by fitting observations with
synthetic spectral profiles using a particular model atmosphere
(temperature, density ans so forth are prescribed), which makes it a
strongly model-dependent process.  Thus, the observations that lead to
the O abundance are not conclusive and arguments exist both in favor
and against the revision. The controversy on whether the proposed
revision should be adopted and the doubts that it would cast on
stellar structure and evolution models is serious enough that it is
often referred to as the solar oxygen crisis (\citeNP{ayresetal06}).




Both the traditional and the proposed revision of the O abundance have
been obtained with similar sets of observations
(an irradiance spectrum and an average disk center spectrum) and
very similar techniques (fitting the equivalent width of atomic
and molecular abundance indicators with synthetic spectra). The main
difference prompting the revision is the model atmosphere employed for
the synthesis. The traditional abundance is obtained when a
semi-empirical one-dimensional (1D) model is used (e.g., that of
\citeNP{holweger-muller74}), whereas the new low abundance is obtained
when using a three-dimensional (3D) hydrodynamical simulation of photospheric
convection (\citeNP{asplundmodel}).



One criticism that has been formulated (\citeNP{ayresetal06}) concerning
the \citeN{asplundetal04} work is that a 3D theoretical model is not
necessarily superior to a 1D semi-empirical one when used to fit
observations. This is a legitimate concern that deserves some
consideration. An attempt to resolve the issue was made by
\citeN{SNN07},
who employed spatially-resolved observations of \FeI \, lines to derive a
3D semi-empirical model (thus combining the advantages of both
strategies) and used it to analyze simultaneous spectra of the \OI \,
infrared triplet at 7770 \AA. In this manner they derived \EO \, at each spatial
position, as opposed to previous single-valued determinations, obtaining
an average \EO=426 ppm. While this result supports the new low
abundace, the authors also pointed out that the spatial distribution
of \EO \, exhibits an unsettling degree of structure, suggesting that
the model is still less than perfect. This is not entirely surprising
since the \OI \, infrared triplet is strongly affected by non-LTE
effects and it is possible that the line formation physics is not yet
completely well understood.

In this work we take a novel approach and analyze Stokes~$V$ 
spectral data of the forbidden [\OI] 6300 \AA \,
line. Its formation physics is much simpler than that of the infrared
triplet, although it is a very weak line and is blended with a \NiI
\, line of similar strength. This blend, rather than being a problem,
turned out to be beneficial for our study because, even though it
complicates the determination of \EO, it provides us with a nearly
model-independent value of the ratio of \EO/\ENi \, as we show below.

\section{Data analysis and results}

The observations employed here were acquired with the
Spectro-Polarimeter for Infrared and Optical Regions (SPINOR,
\citeNP{spinor}) on July 28, 2007. We obtained simultaneous
observations of the \FeI \, lines at 6301.5 and 6302.5~\AA \, and the
6300.3~\AA \, blend of the [\OI] and \NiI \, lines.
The goal of our campaign was to observe a sunspot umbra because the
different Zeeman splitting of the lines would allow us to
differentiate them in the polarized spectrum. Since the Stokes~$V$
profile is normally antisymmetric, the spectral blend gives rise to a
complex structure that provides us with important clues as we explain
below. These lines are very weak and therefore it is important to
observe a strongly magnetized region (ideally a sunspot umbra) to
maximize the polarization signal. Besides, umbral profiles are very
symmetric as velocity gradients are almost nonexistent in the umbral
photosphere (below 100 ms$^{-1}$). Therefore, we can be certain that
any asymmetries we may encounter are due to the presence of the blend.

We observed a small
sunspot (solar activity was very low during our observing run) with
moderate seeing conditions that we estimated to be approximately
1.5\arcsec. Data reduction and polarization calibration were carried
out using the standard procedures for this instrument, which include
correction for the instrumental polarization introduced by the
telescope. We selected the best datasets of July 28: two consecutive
scans 20 minutes appart early in the morning, which we averaged; and
a third scan taken later in the day. We shall refer to these three 
datasets as Maps 1, 2 and 3,
respectively. 

Figure~\ref{fig:profs} shows the average Stokes~$V$ observed in the 
umbra of Map 1. All three
datasets have a very similar average umbral profile. Also depicted in
the figure are synthetic line shapes of the O (blue) and Ni (green)
components of the blend computed separately as well as combined
(red). The model atmosphere used for the synthesis was obtained from
the inversion of the \FeI \, lines.  All syntheses and inversions
presented in this paper were carried out with the LTE code LILIA
(\citeNP{lilia}). Tests were conducted with another standard code (SIR, \citeNP{sir})
using several of the models, to ensure that the same results were obtained.



Notice in Figure~\ref{fig:profs} how the [\OI] and \NiI \, lines have
similar amplitudes. However, the differences in their rest wavelengths
and effective Land\'e factors give rise to a peculiar and strongly
asymmetric shape of the blend when they are combined. 
Particularly interesting is the feature marked by the arrow
in the figure, exactly at the switchover point between the [\OI] and
the \NiI \, blue lobes. This feature is present in all three of our
datasets. 

\begin{figure*}
\plotone{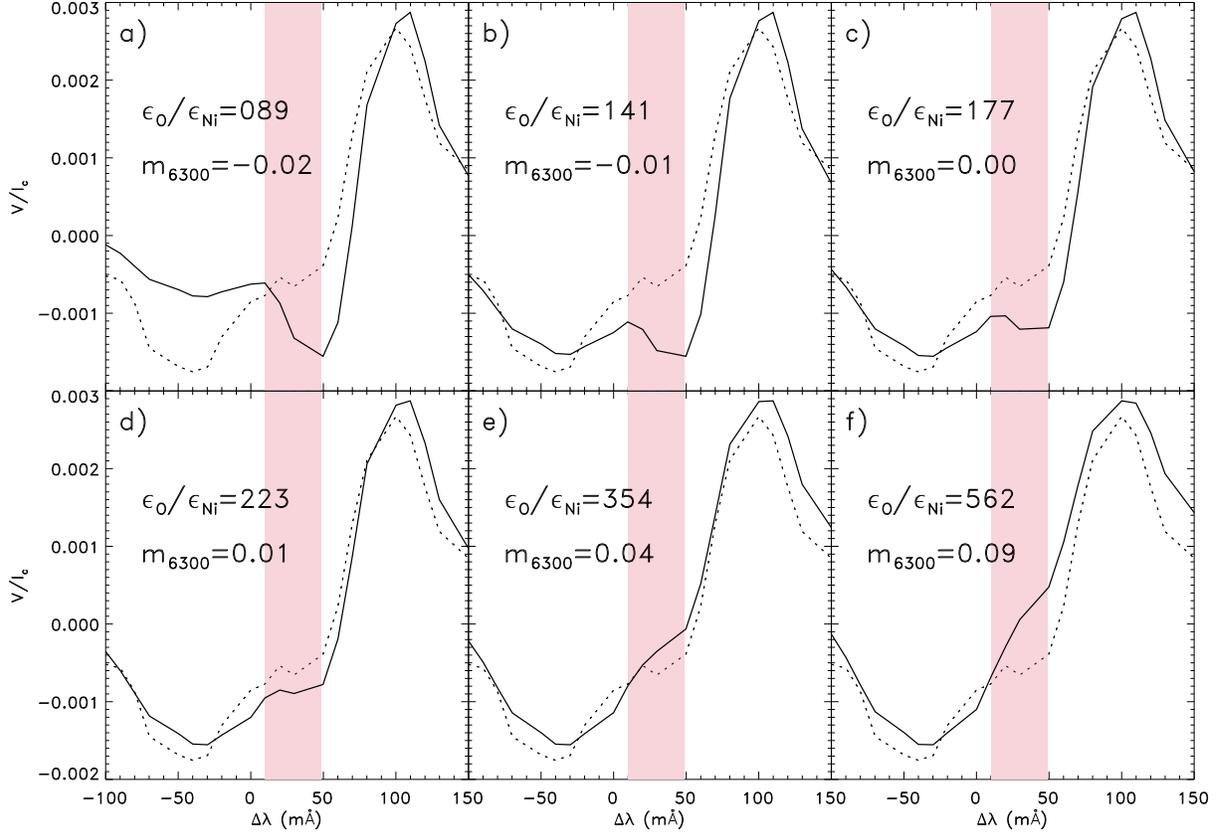}
\caption{Synthetic Stokes~$V$ profile (solid) of the [\OI]/\NiI \,
  blend for different values of the abundance ratio \EO/\ENi . The
  shaded area represents the wavelength range over which we average
  the slope of the profile to obtain $m_{6300}$. For very low ratios
  (panel a), the slope is negative. As the ratio increases (panels
  b, c) $m_{6300}$ becomes less negative until it reaches zero. For
  higher ratios $m_{6300}$ continues to increase (panels d, e, f) until
  the [\OI] line dominates the blend and the profile takes on the normal
  antisymmetric Stokes~$V$ shape of an isolated line.  The
  profile amplitude and zero point depend on details of the model
  employed for the synthesis and the data calibration. For this
  reason, the profiles in the figure have been renormalized to have
  the same amplitude as the observed one (dotted line). However, the
  shape of the central feature in the shaded area and the mean slope
  $m_{6300}$ are virtually model-independent.  Therefore we base our
  analysis on this feature. After correcting for molecular formation,
  and assuming \ENi=1.7, the ratios corresponding to the new and old
  O abundances would be 130 and 200, respectively. The corresponding
  profiles would be similar to those shown in panels b (new abundance)
  and d (old abundance).
\label{fig:samples}
}
\end{figure*}

Consider the average of the slope of the red curve in the shaded area
(hereafter referred to as $m_{6300}$). If we had much more O than Ni,
the red curve would follow the blue curve and this slope would be
positive. Conversely, if we had much more Ni than O it would be the
green curve that dominates and the slope would be negative. In an
intermediate situation, $m_{6300}$ takes values between these two
limits. Figure~\ref{fig:samples} shows this behavior in detail for a
range of the ratio \EO/\ENi \, from low to high values. The
amplitude of the profile and the vertical position of the central
feature depend on the details of the model employed in the synthesis
as well as the polarization calibration. However, the shape of the
feature in the shaded area of Fig~\ref{fig:samples} (and therefore the
mean slope $m_{6300}$) is nearly model-independent, as we show below,
which makes this feature valuable from a diagnostic point of view.

Table~\ref{table:lines} lists the atomic parameters that we
employed. Accurate determinations of the oscillator strength exist for
all lines (\citeNP{Ogf}, \citeNP{Nigf}, \citeNP{Fegf}; see also the
database of the National Institute of Standards and Technology,
NIST\footnote{http://physics.nist.gov/PhysRefData/ASD/index.html}),
except for \FeI \, 6302.5 \AA . We determined the $\log(gf)$ of
this line using the following procedure. We first inverted a
spatially-averaged quiet-Sun profile of \FeI \, 6301.5 \AA . With the
resulting model we synthesized 6302.5 \AA \, varying its $\log(gf)$
until a satisfactory agreement with the observed profile was
attained. We note that the accuracy of the \FeI \, oscillator
strengths is not very important in our work. We use the \FeI \, lines
to determine a suitable model atmosphere. However, the arguments given
below are nearly model-independent. In fact, one could use standard
sunspot models published in the literature and still obtain the same
results as we demonstrate using the model of \citeN{maltbyetal}.

The electric dipole and the magnetic quadrupole components of the [\OI \,] feature 
are listed separately in the table, although the electric dipole contribution is insignificant.
The two \NiI \, components correspond to the two major isotopes 
($^{58}$Ni and $^{60}$Ni): the relative abundances have been folded into the cited 
gf-values. The \OI \, line is well described in LS coupling but
the \NiI \, line exhibits some small departures (see the atomic level
information compilation by NIST). We took this small departure into
consideration in computing the polarized line profile. 

We computed many different models by inverting the \FeI \, lines
observed in the umbra of Maps 1, 2 and 3 (the profiles for Maps 1 and
2 were averaged together to improve the signal-to-noise ratio since
they were taken very close in time) and using different values for the
Fe abundance between 20 and 40 ppm. For each one of these models we
synthesized the Stokes~$V$ spectrum (an example is shown in
Fig~\ref{fig:profs}) and computed $m_{6300}$ as the average slope over
a bandpass of 50~m\AA \, around the [\OI] line center (the vertical
shaded area in Fig~\ref{fig:profs}). When performing all of these
syntheses we added random perturbations to the oscillator strengths of
the lines. The amplitude of these perturbations was set to the
uncertainties in the values published in the literature (14\% and 16\%
for the \NiI \, and \OI \, lines, respectively). In this manner we could
test the impact of these uncertainties on our results. The last column
in Table~\ref{table:lines} lists the range of abundances for the
various elements involved in the synthesis that was used to produce
the figure.

Figure~\ref{fig:asymvsratio} (blue shaded areas) shows how $m_{6300}$
varies as a function of the O/Ni abundance ratio
(\EO/\ENi).  Note that all of these models produce a similar curve of
$m_{6300}$ vs \EO/\ENi, regardless of which dataset we used to derive
the model, what Fe abundance we considered, etc. The spread, delimited
by the shaded region, is due a small degree of model dependence as
well as the random variations in the O and Ni oscillator strengths. As
a further test, we repeated the computation using a standard model
umbra (model M of \citeNP{maltbyetal}). The model temperature was
first normalized to produce the observed continuum intensity
(otherwise the ratio of \OI \, to \NiI \, would be incorrect).
The values of $m_{6300}$ produced with the \citeN{maltbyetal} model
are still inside the band that we obtained with the models from our
inversions. We may then conclude that the
calibration curve of Fig~\ref{fig:asymvsratio} is extremely robust, in
the sense that it exhibits very little sensitivity to the model
atmosphere employed to obtain it.

The horizontal dashed line in Fig~\ref{fig:asymvsratio} marks the
value $m_{6300}$ of the observed profiles (along with the
uncertainty due to observational noise), compatible
with a ratio \EO/\ENi=210$\pm$24. Interestingly, the Ni abundance
currently is better determined than that of O because there are more
well-suited photospheric lines of Ni in the solar
spectrum. \citeN{solarcomposition} quote a current value of
\ENi=1.7$\pm$0.15 ppm, in good agreement with the
meteoritic abundance of \ENi=1.5$\pm$0.1. 

The abundance of {\em atomic} O that we obtain is
\EO=360$\pm$50~ppm.

Since we are observing a sunspot, one must also be aware that some of
the O may be in the form of molecules. Such molecular O does not
contribute to the formation of the 6300~\AA\ blend and has
been neglected thus far. According to \citeN{leemolec}, the most
abundant O molecule (and the only one with a non-negligible
concentration) in sunspot umbrae is CO, which contains nearly 50\% of
all the O. Instead of relying on this value, we made a full
calculation for the temperature of our sunspot. We felt that this was
important because, as noted above, the sunspot that we observed was a
small one and it might be too warm compared to standard models. Our
code solves the molecular chemical equilibrium for an arbitrary number
of molecules.
The interested reader can find a detailed account of the molecular
calculation in a forthcoming research note. Here we shall simply state
that 35 diatomic molecules which potentially can affect the atomic O an Ni
partial pressures were included (\citeN{ST84}). The 6300~\AA\ blend forms nearly at the
same height as the continuum and therefore it is straightforward to
obtain the temperature there. One simply needs to take the continuum
intensity and equate it to the Planck function at that wavelength (as
a sanity check, our inversions of the \FeI\ lines yield approximately
the same temperature at $\tau_{5000}=1$). With this calculation we
obtained that all the molecules containing O and Ni can be neglected
except for CO, which carries about 51\% of the O nuclei at that
temperature. 

It is important to note that the molecular calculation that we have
solved is simply that of chemical equilibrium in LTE (no radiative transfer
or spectral synthesis). The LTE/CE approximation is valid owing to the high 
densities of the umbral deep photosphere. This calculation does not require a model.
Only the gas temperature is needed, which can be extracted directly
from the observations as explained above. Therefore, we are not
introducing any significant complication to our analysis. Putting all
these numbers together we conclude that the total amount of O in the
umbra is \EO=730$\pm$100~ppm, or \lgEO=8.86$\pm$0.07.

\section{Conclusions}

Oxygen is a very important element because of its high abundance and
also because other relevant elements that do not produce suitable
photospheric lines can only be measured relative to it. Unfortunately,
there are few O abundance indicators in the solar spectrum and they
all present complications of one kind or another and require very
detailed modeling to fit the observations. This means that virtually
every abundance determination is susceptible of criticism. Moreover,
it is important to keep in mind that abundance determinations
inherently are model-dependent.

\begin{figure*}
\plotone{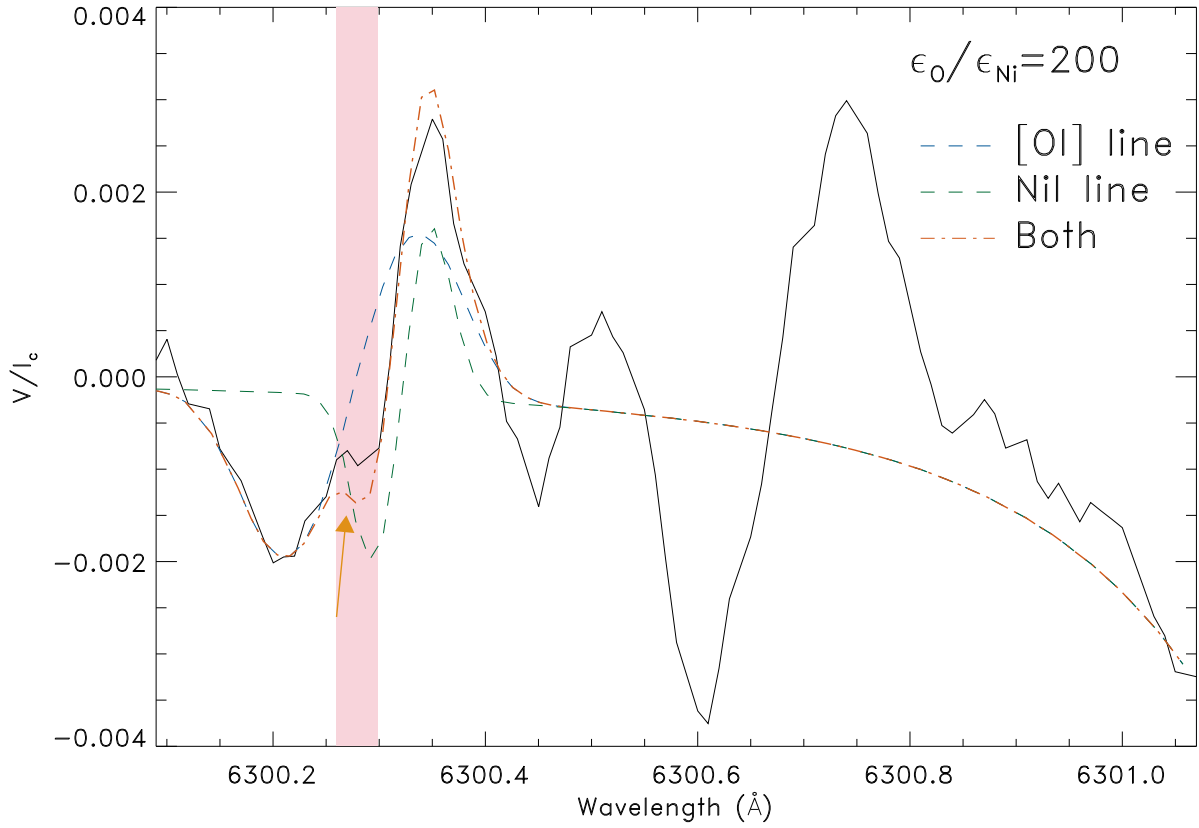}
\caption{Stokes $V$ spectrum of a sunspot umbra. Black,
  solid: observed. Red, dash-dotted: Synthesis of the [\OI] , \NiI
  \, and \FeI \, lines combined. Blue, dashed: Synthesis of [\OI] \, and
  \FeI \, lines 
  only. Green, dashed: Synthesis of \NiI \, and \FeI \, lines only. All
  profiles are normalized to the average disk-center quiet-Sun
  continuum.
\label{fig:profs}
}
\end{figure*}

\begin{figure*}
\plotone{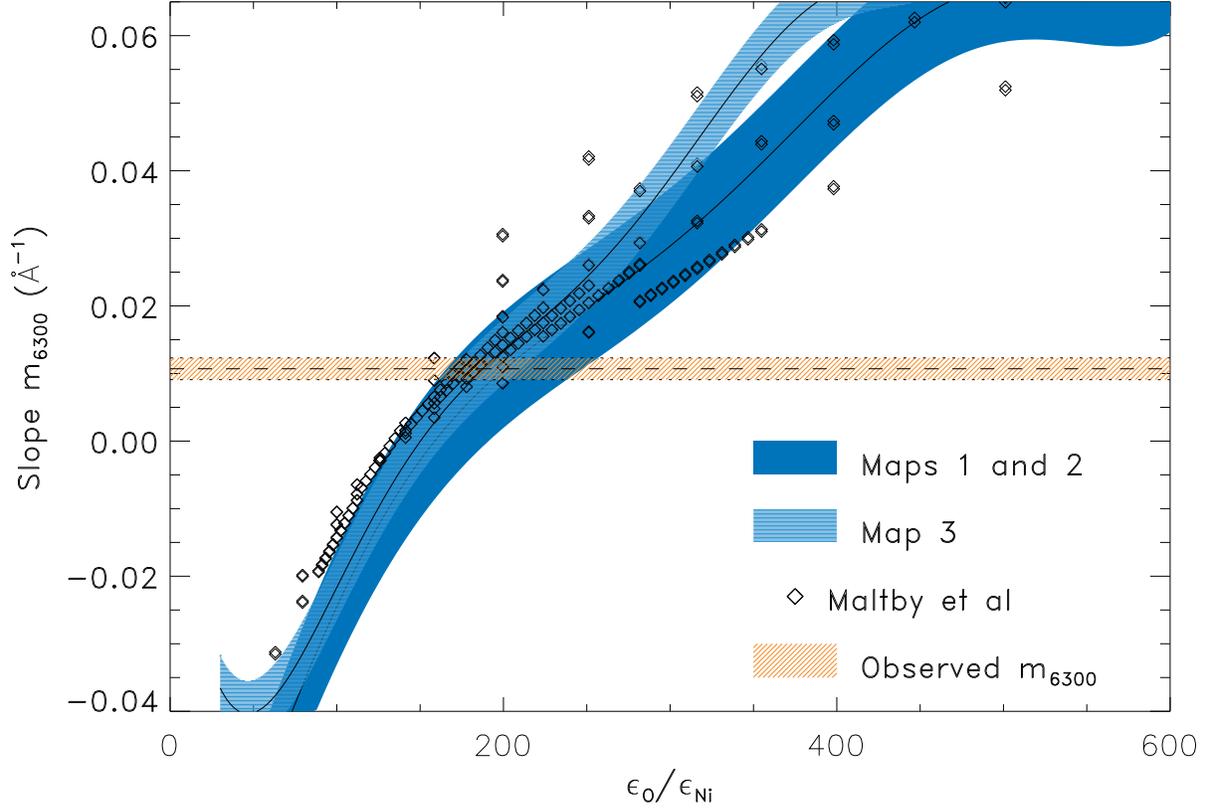}
\caption{Mean slope $m_{6300}$ of the central spectral feature (vertical
  shaded area in Fig~\ref{fig:profs}) as a function of the
  abundance ratio between O and Ni (\EO/\ENi). The solid lines show
  the average curve obtained from many different inversions of the
  \FeI \, lines using different values for the Fe abundance, with
  \EFe \, in the range 20, 40 ppm. The blue shaded areas represent
  the 1-$\sigma$ spread for all of those models. The diamonds
  represent the same curve obtained when a standard sunspot model
  (model M of \citeNP{maltbyetal}) is employed. The horizontal dashed line
  shows the value for $m_{6300}$ determined from the observed
  profile. The orange band indicates the uncertainty of the
  measurement due to noise. All three datasets yield a $m_{6300}$ that
  is within the orange band.
\label{fig:asymvsratio}
}
\end{figure*}

In this work we present a new approach that circumvents these problems
because: (a) we use a spectral feature whose formation physics is well
known (LTE line formation, well-known oscillator strengths); and (b) the
relation between the parameter $m_{6300}$ and the ratio \EO/\ENi \, is
almost model-independent. By analyzing a polarization profile we are
insensitive to common systematic errors, such as flatfield
uncertainties (polarization is a differential measurement). The
polarization profile of the 6300~\AA \, feature is very
peculiar. Instead of analyzing properties such as the amplitude of the
profile, its width or depth, etc, which are sensitive to details of
the modeling, we can focus on its shape. Some features such as
$m_{6300}$ depend very critically on the O and Ni abundances. The fact
that the simplest O abundance indicator (only LTE atomic line in the
visible spectrum) is blended with another line of similar strength has
been viewed historically by many researchers as a rather unfortunate
coincidence. In the light of this new approach, however, one could now
argue the opposite. The abundances of O and Ni in the solar
photosphere are ``just right'' to produce the peculiar polarization
feature marked by the arrow in Fig~\ref{fig:profs}. If either the O or
the Ni line had dominated over the other then we would have had no
other option than fitting the profile amplitude or width, which
again would have been a model-dependent determination and therefore
subject to question.

The ratio \EO/\ENi=210$\pm$24 that we obtain is a very robust result
and this ratio, combined with the well-established Ni abundance and
taking into account that 51\% of the O is in the form of CO, turns out
to be incompatible with the low O abundance proposed by
\citeN{asplundetal04}.

\acknowledgments 
The authors are very grateful to Dr Thomas Ayres for
pointing out the issue of the CO formation that we had missed in our
initial manuscript. Without his suggestion this paper would have
presented very different results.  Thanks are also due to the staff at
the Sacramento Peak observatory (Sunspot, NM, USA) of the National
Solar Observatory for their enthusiastic support of our observing
campaign.  Financial support by the Spanish Ministry of Education and
Science through project AYA2007-63881 is gratefully acknowledged.

\bibliographystyle{../../bib/apj}
\bibliography{ms.bib}

\begin{thebibliography}{18}
\expandafter\ifx\csname natexlab\endcsname\relax\def\natexlab#1{#1}\fi

\bibitem[{{Anders} \& {Grevesse}(1989)}]{anders-grevesse89}
{Anders}, E. \& {Grevesse}, N. 1989, \gca, 53, 197

\bibitem[{{Asplund} {et~al.}(2004){Asplund}, {Grevesse}, {Sauval}, {Allende
  Prieto}, \& {Kiselman}}]{asplundetal04}
{Asplund}, M., {Grevesse}, N., {Sauval}, A.~J., {Allende Prieto}, C., \&
  {Kiselman}, D. 2004, \aap, 417, 751

\bibitem[{{Asplund} {et~al.}(1999){Asplund}, {Nordlund}, {Trampedach}, \&
  {Stein}}]{asplundmodel}
{Asplund}, M., {Nordlund}, {\AA}., {Trampedach}, R., \& {Stein}, R.~F. 1999,
  \aap, 346, L17

\bibitem[{{Ayres} {et~al.}(2006){Ayres}, {Plymate}, \& {Keller}}]{ayresetal06}
{Ayres}, T.~R., {Plymate}, C., \& {Keller}, C.~U. 2006, \apjs, 165, 618

\bibitem[{{Bard} {et~al.}(1991){Bard}, {Kock}, \& {Kock}}]{Fegf}
{Bard}, A., {Kock}, A., \& {Kock}, M. 1991, \aap, 248, 315

\bibitem[{{Basu} \& {Antia}(2008)}]{basu-antia2008}
{Basu}, S. \& {Antia}, H.~M. 2008, Physics Reports, 457, 217

\bibitem[{{Frose-Fischer} \& {Saha}(1983)}]{Ogf}
{Frose-Fischer}, C.~F. \& {Saha}, H.~P. 1983, \pra, 28, 3169

\bibitem[{{Grevesse} {et~al.}(2007){Grevesse}, {Asplund}, \&
  {Sauval}}]{solarcomposition}
{Grevesse}, N., {Asplund}, M., \& {Sauval}, A.~J. 2007, Space Science Reviews,
  130, 105

\bibitem[{{Grevesse} \& {Sauval}(1998)}]{grevesse-sauval98}
{Grevesse}, N. \& {Sauval}, A.~J. 1998, Space Science Reviews, 85, 161

\bibitem[{{Holweger} \& {Mueller}(1974)}]{holweger-muller74}
{Holweger}, H. \& {Mueller}, E.~A. 1974, \solphys, 39, 19

\bibitem[{{Johansson} {et~al.}(2003){Johansson}, {Litz{\'e}n}, {Lundberg}, \&
  {Zhang}}]{Nigf}
{Johansson}, S., {Litz{\'e}n}, U., {Lundberg}, H., \& {Zhang}, Z. 2003, \apjl,
  584, L107

\bibitem[{{Lee} {et~al.}(1981){Lee}, {Kim}, {Yun}, {Beebe}, \&
  {Davis}}]{leemolec}
{Lee}, H.~M., {Kim}, D.~W., {Yun}, H.~S., {Beebe}, R., \& {Davis}, R. 1981,
  Journal of Korean Astronomical Society, 14, 19

\bibitem[{{Maltby} {et~al.}(1986){Maltby}, {Avrett}, {Carlsson},
  {Kjeldseth-Moe}, {Kurucz}, \& {Loeser}}]{maltbyetal}
{Maltby}, P., {Avrett}, E.~H., {Carlsson}, M., {Kjeldseth-Moe}, O., {Kurucz},
  R.~L., \& {Loeser}, R. 1986, \apj, 306, 284

\bibitem[{{Ruiz Cobo} \& {del Toro Iniesta}(1992)}]{sir}
{Ruiz Cobo}, B. \& {del Toro Iniesta}, J.~C. 1992, \apj, 398, 375

\bibitem[{{Sauval} \& {Tatum}(1984)}]{ST84}
{Sauval}, A.~J. \& {Tatum}, J.~B. 1984, \apjs, 56, 193

\bibitem[{{Socas-Navarro}(2001)}]{lilia}
{Socas-Navarro}, H. 2001, in Astronomical Society of the Pacific Conference
  Series, Vol. 236, Advanced Solar Polarimetry -- Theory, Observation, and
  Instrumentation, ed. M.~{Sigwarth}, 487--+

\bibitem[{{Socas-Navarro} {et~al.}(2006){Socas-Navarro}, {Elmore}, {Pietarila},
  {Darnell}, {Lites}, {Tomczyk}, \& {Hegwer}}]{spinor}
{Socas-Navarro}, H., {Elmore}, D., {Pietarila}, A., {Darnell}, A., {Lites},
  B.~W., {Tomczyk}, S., \& {Hegwer}, S. 2006, \solphys, 235, 55

\bibitem[{{Socas-Navarro} \& {Norton}(2007)}]{SNN07}
{Socas-Navarro}, H. \& {Norton}, A.~A. 2007, \apjl, 660, L153

\end{thebibliography}

\clearpage

\begin{deluxetable}{cccccccc}
\tablewidth{0pt}
\tablecaption{Atomic parameters
\label{table:lines}}
\tablehead { Ion &Wavelength & Excitation & $\log(gf)$ & Configuration &
  Configuration & Land\'e & Abundance \\
             &   (\AA)     & potential (eV) &        & (lower)  &
  (upper) & factor & range (ppm) }
\startdata
   \OI & 6300.304 &  0   &    -9.776  &  $^3$P$_2$ & $^1$D$_2$ & 1.25
   & 200-630 \\
   \OI & 6300.304 &  0   &    -12.202  &  $^3$P$_2$ & $^1$D$_2$ & 1.25
   & 200-630 \\
   \NiI & 6300.335 & 4.266 & -2.250 & $^3$D$_1$ & $^3$P$_0$ & 0.51 & 1-3 \\
   \NiI & 6300.355 & 4.266 & -2.670 & $^3$D$_1$ & $^3$P$_0$ & 0.51 &
   1-3 \\
   \FeI & 6301.501 & 3.654 & -0.718 & $^5$P$_2$ & $^5$D$_2$ & 1.67 & 20-40\\
   \FeI & 6302.492 & 3.686 & -1.235 & $^5$P$_1$ & $^5$D$_0$ & 2.50 & 20-40\\
\enddata
\end{deluxetable}

\clearpage

\end{document}